\documentclass[article]{IEEEtran}
\usepackage{graphicx} 
\usepackage{amsmath}
\usepackage{amssymb}
\usepackage{physics}
\usepackage{cite}
\usepackage{tikz}
\usepackage[T1]{fontenc}
\usepackage{multirow}

\newcommand{\M}[1]{\mathbf{#1}}
\newcommand{\T}[1]{\mathrm{#1}}
\newcommand{\V}[1]{\boldsymbol{#1}}
\renewcommand{\u}[1]{\hat{\boldsymbol{#1}}}

\newcommand{\sens}{\chi}

\title{Directivity Estimates and Tolerance Tradeoffs in the Presence of Random Perturbations}

\author{Kurt Schab, \IEEEmembership{Member, IEEE}, Lukas Jelinek, Miloslav Capek, \IEEEmembership{Senior Member, IEEE}, \\and Mats Gustafsson, \IEEEmembership{Senior Member, IEEE}
\thanks{Manuscript received \today; revised \today. \textit{(Corresponding author: K. Schab.)}}
\thanks{This work was supported by the Czech Science Foundation under project No. 24-11678S, and Swedish Research Council SEE-6GIA 2024-06482.}
\thanks{K. Schab is with Santa Clara University, Santa Clara, USA (e-mail: kschab@scu.edu).}
\thanks{L. Jelinek and M. Capek are with Czech Technical University in Prague, Czech Republic (e-mail: lukas.jelinek@fel.cvut.cz, miloslav.capek@fel.cvut.cz).}
\thanks{M. Gustafsson is with Lund University, Lund, Sweden, (e-mail: mats.gustafsson@eit.lth.se).}
}

\begin{document}


\pagestyle{headings}
\twocolumn

\maketitle
\begin{abstract}
    The practicality of realizing high-directivity sources is studied using a generalized sensitivity metrics.  These measures are based on the error model adopted in previous studies on array tolerance, but account for heterogeneous array characteristics in an arbitrary source-expansion basis, making it applicable to the study of continuous sources, ideal uncoupled arrays, and heterogeneous arrays with mutual coupling.  The formulation provides the ensemble average of realized directivity under the selected error model, making it a fast predictor of anticipated directivity reduction due to source perturbation.  The data required to compute these predictions are straightforward to obtain via simulation or measurement, with a significant simplification when the system can be approximated as lossless.  The directivity and the proposed tolerance measures allow for the formulation of optimization problems illustrating the tradeoff between these measures.  Several examples demonstrate these tradeoffs, special cases, comparison to previous estimates of maximum directivity, and the accuracy of sensitivity as a predictor of statistical performance over large ensembles of design realizations.
\end{abstract}


\section{Introduction}

High directivity sources, whether arrays or single elements, have long been known to suffer from high sensitivity to manufacturing, excitation, or placement errors~\cite{cheng2005optimization,1981_Hansen_PIEEE,diao2017practical,william2025comparative,krasnok2026advanced}.  Seminal work on this topic explored the notion of statistical predictors for changes in directive radiation characteristics under various perturbations to ideal, uncoupled arrays and deviations from ideal aperture source distributions~\cite{ruze1952physical,gilbert1955optimum,1956_Uzsoky_APASH,elliott1958mechanical,rondinelli1966effects,richards1975antenna}.  The same line of reasoning was applied by various authors in the following years\footnote{See Collin and Zucker, \textit{Antenna Theory Vol. 1}, 1969 for a comprehensive bibliography of work from this era.}, eventually giving rise to the widely-adopted concept of ``array Q-factor''~\cite{1956_Uzsoky_APASH,lo1966optimization,pozar1984antenna} as a measure of pattern sensitivity in the presence of certain classes of errors.  Further work utilized a variety of techniques to derive predictive statistics for entire patterns~\cite{howard1967statistical,sjoberg2005coherent,lee2021accurate}, with many authors focused on the assumption of homogeneous arrays with isotropic elements~\cite{rondinelli1966effects,hsiao1985design,anselmi2013tolerance,moghaddam2019statistical,trucco2022statistics}.  While these prior approaches clearly illustrate the general concept of a directivity-sensitivity tradeoff, they are based on simplifying assumptions that neglect mutual coupling or assume homogeneous element patterns, making it inapplicable to many systems of practical interest.  To study such problems in the absence of statistical predictors, Monte Carlo analysis methods are required~\cite{yaghjian2008electrically,yen2025practicality}, which provide a ground truth of any error statistics derived by other means.  Similarly, Monte Carlo methods have been applied to the problem of determining error tolerances of individual elements within ideal arrays~\cite{lee2005decision}.

In this work, we generalize the concept of array tolerance to the analysis of arbitrary sources, including continuous current distributions, modal expansions, and arrays of coupled antennas with heterogeneous embedded element patterns.  Two metrics are formed, uniformity and sensitivity, which predict directivity perturbation under general error models.  These metrics are computed using embedded element pattern data and, in the case of lossless systems, can be partially represented by network scattering parameters.  We then study the structure of these predictors under specific assumptions regarding error correlation and scaling.  By formulating a multi-objective optimization problem between directivity and sensitivity, we obtain quantitative tradeoffs between main-beam directivity and its mean reduction due to errors.  Numerous examples study the accuracy of the derived predictors against Monte Carlo analysis, the tradeoff between sensitivity and directivity in superdirective endfire arrays, including mutual coupling, scaling with electrical size, and the method's application to continuous current distributions.

\section{Radiation Intensity from Perturbed Sources}
In this section, we formulate the variation in directivity due to perturbations to a prescribed source.  Without loss of generality, the formulation considers ``sources'' by current distributions $\V{J}$ confined to a region $\varOmega$.   The $\u{\gamma}^*$-polarized far field ($^*$ denoting complex conjugation) produced in the $\u{k}$ direction by this current is given by
\begin{equation}
F_\gamma(\u{k}) = \lim_{r\rightarrow\infty} r\u{\gamma}\cdot\V{E}\T{e}^{\T{j}kr}= \frac{-\T{j}k\eta}{4\pi}\int \u{\gamma}\cdot\V{J}(\V{r})\T{e}^{\T{j}\V{k}\cdot\V{r}}\,\T{d}\V{r},
\label{eq:f-from-j}
\end{equation}
where $\V{E}$ is the electric field produced by the current density $\V{J}$, \mbox{$k = |\V{k}|$} is the freespace wavenumber, $\V{k}=\u{k}k$ is the wavevector, and $\eta$ is the background impedance.  The associated partial radiation intensity in this direction and polarization is
\begin{equation}
    U^\gamma = \frac{|F_\gamma|^2}{2\eta}
    \label{eq:uf2}
\end{equation}
and the partial directivity in the $\u{k}_0$ direction reads
\begin{equation}
    D = \frac{4\pi U^\gamma(\u{k}_0)}{\int U(\u{k})\,\T{d\Omega}}
    \label{eq:d-partial-def}
\end{equation}
where integration is taken over the entire solid angle and 
\begin{equation}
    U = \sum \limits_\gamma U^\gamma
\end{equation}
is the total radiation intensity from both polarizations.  Throughout the following development, we consider partial directivity $D$ only in an intended polarization $\u{\gamma}$ as written in \eqref{eq:d-partial-def}.  Similarly, the term ``radiation intensity'' will generally refer to partial radiation intensity $U^\gamma$ unless explicitly stated otherwise. The small modification in notation to cast all results in terms of the total directivity over both polarizations is discussed in Sec.~\ref{sec:directivity-estimates}.

\begin{figure}
    \centering
    \includegraphics[width=\linewidth]{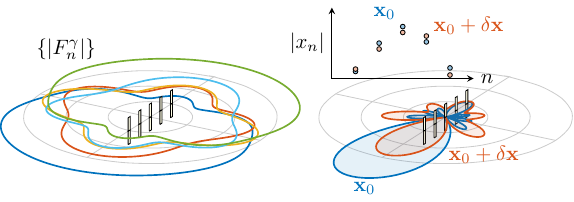}
    \caption{(Left) Embedded element patterns $|F_n^\gamma|$ of a five element array.  (Right) Total pattern $F^\gamma$ for a nominal endfire excitation $\M{x}_0$ and a perturbed excitation $\M{x}_0+\delta\M{x}$.}
    \label{fig:setup-array}
\end{figure}

\begin{figure}
    \centering    
    \includegraphics[width=\columnwidth]{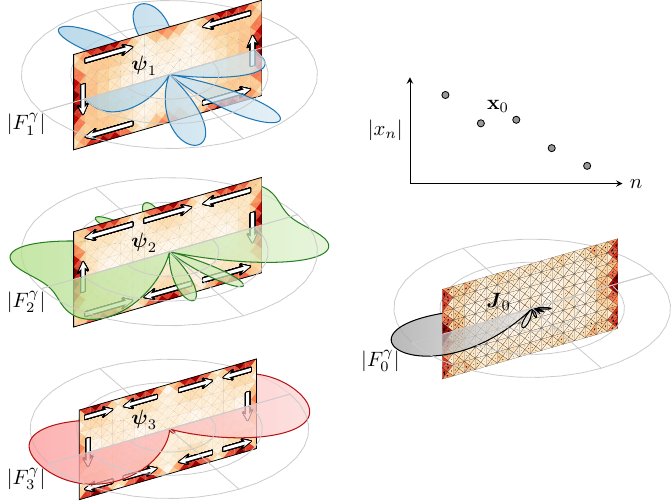}
    \caption{(Left) Modal currents $\V{\psi}_n$ and radiation patterns $|F_n^\gamma|$ as a basis for describing radiation from currents confined to a rectangular plate.  (Right) Representation of a general current $\V{J}_0$ and its radiation pattern $|F_0^\gamma|$ for endfire radiation in terms of the selected basis.  This current is constructed by the sum of modal currents with coefficients $\M{x}_0$.  It is complex-valued and only the amplitude is shown.}
    \label{fig:setup-modes}
\end{figure}

\subsection{Error Model}
To consider how the directivity in \eqref{eq:d-partial-def} varies under small perturbations of the source current, let the total current distribution be the sum of a nominal current $\V{J}_0$ and a random error vector $\delta\V{J}$, i.e.,
\begin{equation}
    \V{J} = \V{J}_0 + \delta\V{J}.
    \label{eq:err-1}
\end{equation}
We further consider the current to be represented in a basis~$\V{\psi}_n$ such that
\begin{equation}
    \V{J} = \sum_n(x_{0n} + \delta x_n)\V{\psi}_n.
    \label{eq:j-basis}
\end{equation}
The nominal coefficients~$x_{0n}$ and errors~$\delta x_n$, collected in the column vectors $\M{x}_0$ and $\delta\M{x}$, respectively, fully define a perturbed current.  The vector $\delta\M{x}$ is a random variable described by its statistics.  For generality, we make no initial assumptions regarding the mean or correlation properties of the errors $\delta\M{x}$, though the special case of zero-mean, uncorrelated errors is common throughout the literature and studied extensively in later sections. 

\subsection{Expected Value of Radiation Intensity}
To infer the effect of the random variations $\delta\M{x}$ on the directivity produced by the nominal coefficients $\M{x}_0$, we begin by examining the expected value of the radiation intensity following the procedure used in~\cite{gilbert1955optimum} for ideal uncoupled arrays.  In the selected basis, the $\u{\gamma}^*$-polarized far field in the $\u{k}$ direction is given by 
\begin{equation}
    F^\gamma = \M{F}^\gamma\M{x},
\end{equation}
where the row vector $\M{F}^\gamma$ collects entries
\begin{equation}
    F^\gamma_n = -\frac{\T{j}k\eta}{4\pi}\int \u{\gamma}\cdot\V{\psi}_n \T{e}^{\T{j}\V{k}\cdot\V{r}}\T{d}\V{r}
    \label{eq:fm}
\end{equation}
which is the far-field produced by the $n$-th basis function~$\V{\psi}_n$.  The far field $F_n^\gamma$ can have different interpretations depending on the choice of the basis $\{\V{\psi}_n\}$. If the total current $\V{J}$ represents the driven solution on an array with some number of ports, then the basis functions~$\V{\psi}_n$ are current profiles each excited by the individual port (with or without mutual coupling) and the terms $F_n^\gamma$ are the embedded element patterns of the array, see the left panel of Fig.~\ref{fig:setup-array}.  Alternatively, if the current is expanded into global basis functions (e.g., characteristic modes~\cite{HarringtonMautz1971a}), then $F_n^\gamma$ represent the modal radiation patterns, see left panel of Fig.~\ref{fig:setup-modes}.  Problems involving both of these cases are studied in detail in Sec.~\ref{sec:examples}.

Following~\eqref{eq:uf2}, the radiation intensity reads
\begin{equation}
    U^\gamma = \frac{1}{2\eta}|\M{F}^\gamma\M{x}|^2 = \frac{1}{2}\M{x}^\T{H}\M{U}^\gamma\M{x}.
    \label{eq:u-op}
\end{equation}
Given that~$\M{x} = \M{x}_0 + \delta\M{x}$, the radiation intensity is expanded
\begin{equation}
    U^\gamma = \dfrac{1}{2}\M{x}_0^\T{H}\M{U}^\gamma\M{x}_0 + \T{Re} \left\{ \M{x}_0^\T{H}\M{U}^\gamma\delta\M{x} \right\} + \dfrac{1}{2} \delta\M{x}^\T{H}\M{U}^\gamma\delta\M{x},
\end{equation}
where the first term is the nominal radiation intensity of the unperturbed current, denoted in subsequent expressions by $U_0^\gamma$, and $\M{M}^\T{H}$ denotes the Hermitian transpose of matrix~$\M{M}$.  Taking the expected value~$\langle \cdot\rangle$ and applying the cyclic trace property $\T{Tr}(\delta\M{x}^\T{H}\M{U}^\gamma\delta\M{x}) = \T{Tr}(\M{U}^\gamma\delta\M{x}\delta\M{x}^\T{H})$~\cite[(4.30)]{gray2004introduction}, we obtain
\begin{equation}
    \langle U^\gamma\rangle =U_0^\gamma + \T{Re} \left\{ \M{x}_0^\T{H}\M{U}^\gamma\V{\mu} \right\} +  \dfrac{1}{2}\T{Tr}\,(\M{U}^\gamma\M{A}),
    \label{eq:f2-alt}
\end{equation}
where the entries of the column vector~$\V{\mu}$ are the expected values of individual $\delta x_n$ terms and $\M{A}$ is the autocorrelation matrix~\cite[\S4.11]{gray2004introduction}
\begin{equation}
    \M{A} = \langle\delta\M{x}\delta\M{x}^\T{H}\rangle.
\end{equation}
The expected value of the radiation intensity
\begin{equation}
    \langle U^\gamma\rangle = U_0^\gamma + \langle U_\T{b}^\gamma\rangle
    \label{eq:u0-ub}
\end{equation}
is therefore given by the sum of the nominal pattern $U_0^\gamma$ of the unperturbed excitation coefficients $\M{x}_0$ and the background pattern~$\langle U_\T{b}^\gamma\rangle$  depending on the mean and correlation properties of the error terms $\delta\M{x}$.

\subsection{Directivity Estimates}
\label{sec:directivity-estimates}
We now derive statistical predictors for main-beam and sidelobe directivity.  Owing to the generality of \eqref{eq:f2-alt}, these predictors apply to systems with heterogeneous element patterns, mutual coupling, arbitrary distributed element basis functions, and arbitrary error coupling.  General expressions are derived here before studying the implications of several simplifying assumptions.

From \eqref{eq:d-partial-def}, the expected value of the directivity reads
\begin{equation}
    \langle D\rangle = \left\langle\frac{4 \pi U^\gamma}{\int U \,\T{d}\Omega}\right\rangle.
\end{equation}
If the numerator and denominator are reasonably independent~\cite{gray2004introduction} and the variance of the denominator is small relative to its absolute value\footnote{This is most easily seen by taking the mean of a two-dimensional Taylor expansion of a function $f = x/y$ where $x$ and $y$ are independent random variables.  It follows that $\langle f\rangle \approx \langle x \rangle/\langle y\rangle$ when the standard deviation of the denominator $\sigma_y$ is much smaller than its mean, i.e., $\sigma_y\ll\langle y\rangle$.}, the expected value can be approximated by~\cite{gilbert1955optimum}
\begin{equation}
    \langle D\rangle \approx \bar{D} = \frac{4\pi (U^\gamma_0+\left\langle U^\gamma_\T{b}\right\rangle)}{\int (U_0+\left\langle U_\T{b}\right\rangle)\,\T{d}\Omega},
    \label{eq:anticipD1}
\end{equation}
where $\bar{D}$ is the \textit{anticipated} directivity, an estimate of the ensemble mean directivity under a selected error model. The validity of this approximation is tested extensively in subsequent examples.  Substituting \eqref{eq:f2-alt} and \eqref{eq:u0-ub} into~\eqref{eq:anticipD1} leads to 
\begin{equation}
    \bar{D} = \frac{D_0 + \hat{\zeta}}{1+\hat{\chi}},
    \label{eq:dbar-coupled}
\end{equation}
where 
\begin{equation}
    \hat{\zeta} = 4\pi \frac{ \left\langle U^\gamma_\T{b}\right\rangle }{\int U_0\,\T{d}\Omega} = 4\pi\frac{2\T{Re} \left\{ \M{x}_0^\T{H}\M{U}^\gamma \V{\mu} \right\} +  \T{Tr}\,(\M{U}^\gamma\M{A})}{\M{x}_0^\T{H}\M{R}\M{x}_0}
\end{equation}
and
\begin{equation}
    \hat{\chi} = \frac{\int \left\langle U_\T{b}\right\rangle\,\T{d}\Omega}{\int U_0\,\T{d}\Omega} = \frac{2\T{Re} \left\{ \M{x}_0^\T{H} \M{R} \V{\mu} \right\} +  \T{Tr}\,(\M{R}\M{A}) }{\M{x}_0^\T{H}\M{R}\M{x}_0},
\end{equation}
with~$\M{R}$ denoting radiated power matrix
\begin{equation}
    P_\T{rad}  = \frac{1}2{}\M{x}^\T{H}\left(\sum_\gamma\int\M{U}^\gamma\,\T{d}\Omega\right)\M{x} = \frac{1}{2}\M{x}^\T{H}\M{R}\M{x}.
    \label{eq:uint-r}
\end{equation}
The nominal directivity $D_0$ is the directivity
\begin{equation}
    D = 4\pi\frac{\M{x}^\T{H}\M{U}^\gamma\M{x}}{\M{x}^\T{H}\M{R}\M{x}}.
    \label{eq:Dmom}
\end{equation}
evaluated for the nominal excitation $\M{x}_0$.

Computation of the matrices~$\M{R}$ and~$\M{U}^\gamma$ is straightforward using any solver or measurement system capable of producing the far field $F_n^\gamma$ due to each excitation at a variety of observation angles\footnote{The integration in \eqref{eq:uint-r} can be approximated by an appropriate quadrature rule, e.g., that of Lebedev~\cite{LebedevLaikov_QuadratureRuleSphere}.}.  Additionally, significant simplifications can be made in the case of lossless (or approximately lossless) systems, allowing for the computation of the matrix $\M{R}$ from S-parameter data, see Appendix~\ref{app:lossless}.

The relation~\eqref{eq:dbar-coupled} is a predictor of the anticipated directivity~$\bar{D}$ based on the nominal excitation vector~$\M{x}_0$ and the error statistics~$\V{\mu}$ and~$\M{A}$.  The parameters~$\hat{\zeta}$ and~$\hat{\chi}$ represent error coupling into the observation direction and total radiated power, respectively. Assuming small perturbations, the parameter~$\hat{\zeta}$ tends to increase the anticipated directivity when the nominal directivity~$D_0$ is small, e.g., increasing sidelobe and cross-polarization levels. The parameter $\hat{\chi}$ decreases the anticipated directivity in cases of high nominal directivity~$D_0$, i.e., decreases the directivity of the main beam.   

\subsection{Zero-mean Uncorrelated Proportional Errors}
\label{sec:zero-mean}

While \eqref{eq:dbar-coupled} applies to systems with arbitrary error statistics, the inner structure and interpretation of the terms $\hat{\zeta}$ and $\hat{\chi}$ are complicated by the possible non-zero mean terms $\V{\mu}$ and correlation between errors represented by off-diagonal terms in $\M{A}$.  Here we study the commonly-assumed~\cite{bach1969uniform} case of zero-mean~$\V{\mu} = \M{0}$, uncorrelated errors that have variance proportional to the nominal terms they modify, i.e.,
\begin{equation}
    A_{nn} = \langle|\delta x_n|^2\rangle = \epsilon|x_{0n}|^2
    \label{eq:xvarprop}
\end{equation}
and~$A_{mn} = 0$ for~$m \neq n$.  This choice of error model captures the effects of proportional amplitude errors and phase shift quantization errors.  Electrically small mechanical displacements can also be represented by this error model, with the additional assumption that embedded element patterns are unaffected by those displacements.  Representative values of the error scaling parameter $\epsilon$ are shown in Tab.~\ref{tab:error-numbers}.

\renewcommand{\arraystretch}{1.3}
\begin{table}[]
    \centering
    \begin{tabular}{c|c|c}
        \textbf{error source} & \textbf{error amplitude} & $\epsilon$ \\\hline
        \multirow{ 2}{*}{amplitude error} & $\pm 0.1$~dB ($1.2\%$) & $4.5\times10^{-5}$\\
        & $\pm 1$~dB ($12.2\%$) & $5.0\times10^{-3}$\\\hline
        \multirow{2}{*}{phase shift quantization} & 6-bit resolution ($\pm 2.8^\circ$) & $8.0\times10^{-4}$\\
         & 4-bit resolution ($\pm 11.25^\circ$) & $1.3\times10^{-2}$\\\hline
        element displacement & $\lambda/100$ & $1.3\times10^{-3}$\\
       (no mutual coupling) & $\lambda/20$ & $3.3\times10^{-2}$\\\hline
    \end{tabular}\\
    ~\\
    \caption{Example values of error scaling parameter $\epsilon$ for various sources of error.  Amplitude and phase variations are assumed to be uniformly distributed.  Displacement errors are normally distributed in three dimensions with standard deviation given by the error amplitude (see~\mbox{\cite[\S 5]{gilbert1955optimum}.})}
    \label{tab:error-numbers}
\end{table}

Under these assumptions, the anticipated directivity reads
\begin{equation}
    \bar{D} =  \frac{D_0 + \epsilon\zeta}{1+\epsilon\chi},
    \label{eq:dbar-complete}
\end{equation}
where 
\begin{equation}
    \zeta = \frac{4\pi}{\epsilon}\frac{\langle U^\gamma_\T{b}(\u{k}_0)\rangle}{\int U_0(\u{k}) \T{d}
    \Omega} = 4\pi \frac{\M{x}_0^\T{H}(\T{diag} \, \M{U}^\gamma)\M{x}_0}{\M{x}_0^\T{H}\M{R}\M{x}_0},
    \label{eq:zetamom}
\end{equation}
and
\begin{equation}
\sens = \frac{1}{\epsilon}\frac{\int\langle U_\T{b}\rangle(\u{k})\,\T{d}\Omega}{\int U_0(\u{k})\,\T{d}\Omega} 
= \frac{\M{x}_0^\T{H}(\T{diag} \, \M{R})\M{x}_0}{\M{x}_0^\T{H}\M{R}\M{x}_0}.
\label{eq:qsbar}
\end{equation}
are parameters that are independent of the variance scaling $\epsilon$. The term~$\T{diag} \, \M{X}$ represents a diagonal matrix containing only the diagonal entries of a matrix $\M{X}$.  Compare~\eqref{eq:dbar-complete} to \cite[(13b)]{howard1967statistical} for the specific case of zero-mean Gaussian errors with possible non-zero error correlation.

The \textit{uniformity} parameter $\zeta$ represents the weighted sum of radiation intensities in the observation direction $\u{k}_0$ from all uncorrelated degrees of freedom in isolation (numerator) relative to the total radiated power (denominator), normalized by the error variance $\epsilon$.  The \textit{sensitivity} parameter $\sens$ is the uniformity integrated over all observation angles and represents the total powers radiated by individual degrees of freedom relative to the total power of the complete system.  Together in \eqref{eq:dbar-complete}, these two parameters predict deviations in the anticipated directivity $\bar{D}$ away from the nominal directivity pattern $D_0$.  The use of the diagonal matrix construction in the definitions of sensitivity $\chi$ and uniformity $\zeta$ allows for basis-dependent behavior even under a unitary change of basis, a feature explored thoroughly in subsequent examples.

Two simplifications of \eqref{eq:dbar-complete} can be made in practice.  In the high-directivity case when $U_0(\u{k}_0)\gg\langle U_\T{b}(\u{k}_0)\rangle$ (e.g., ``main-beam'' characteristics), the directivity estimate reduces to
\begin{equation}
     \bar{D} \approx \frac{4\pi U_0(\u{k}_0)}{\int (U_0+\left\langle U_\T{b}\right\rangle)\,\T{d}\Omega}  = \frac{D_0}{1+\epsilon\chi}.
     \label{eq:dbar}
\end{equation}
This implies that perturbation-driven reduction in main-beam directivity can be predicted using the sensitivity factor $\chi$ alone.  In contrast, in the low-directivity regime (i.e., near nulls in the nominal pattern where $D_0\approx 0$ or in cases where $D_0$ represents low cross-polarization) the directivity estimate becomes
\begin{equation}
    \bar{D} \approx \frac{4\pi\langle U_\T{b}(\u{k})\rangle}{\int_\Omega U_0(\u{k}) + \langle U_\T{b}(\u{k})\rangle  \,\T{d}\Omega} = \frac{\epsilon\zeta}{1+\epsilon\chi}.
    \label{eq:dbar-null}
\end{equation}
Note that the parameter $\zeta$ depends on observation angle while~$\chi$ does not.  Together, $\zeta$ and $\chi$ predict changes in null depth.  

For the remainder of this paper, we mainly consider the special case of \eqref{eq:dbar}, where deviations in main-beam directivity are quantified solely by the sensitivity $\chi$.  Numerous examples in Sec.~\ref{sec:examples} demonstrate the applicability of that estimate in a variety of problems.  For completeness, Sec.~\ref{sec:estimate-accuracy} contains a detailed study of the ensemble average directivity~$\langle D \rangle$ obtained from Monte Carlo analysis, the general estimate in \eqref{eq:dbar-complete}, and the special cases in \eqref{eq:dbar} and \eqref{eq:dbar-null}.

\subsection{Simplification for Uniform, Ideal, and Uncoupled Arrays}
The general expression in \eqref{eq:f2-alt} for the expected value of radiation intensity simplifies in the special case of uniform, ideal, and uncoupled arrays under perturbed excitations or element positions.  This special case has been studied extensively in the literature~\cite{gilbert1955optimum} and forms the basis for many prior works on sensitivity in array systems~\cite{bach1969uniform,skolnik1969nonuniform}.  When all array elements are assumed to have identical element patterns~$F_\T{e}^\gamma(\u{k})$, the field due to each element takes the form
\begin{equation}
    F_n^\gamma =F_\T{e}^\gamma(\u{k})\T{e}^{\T{j}\V{k}\cdot\V{r}_n},
    \label{eq:element-pattern}
\end{equation}
and the sensitivity and uniformity parameters reduce to 
\begin{equation}
    \chi = D_\T{e}^{-1}\zeta = D_0K,
\end{equation}
where
\begin{equation}
    K = \frac{\sum_n |x_{n0}|^2}{\left|\sum_n x_{n0}\T{e}^{\T{j}\V{k}_0\cdot\V{r}_n}\right|^2}
    \label{eq:k-def}
\end{equation}
is a measure scaling the background pattern that is independent of observation angle~$\u{k}$ and $D_\T{e}$ is the directivity associated with the element pattern $F_\T{e}$.  The factor $K$ is closely related to the sensitivity $S$ used in~\cite{lo1966optimization} and the ``white noise gain'' $G_\T{w}$ studied in~\cite{cox1986practical}.  The anticipated directivity in~\eqref{eq:dbar-complete} reduces to 
\begin{equation}
    \bar{D} = \frac{D_0 + \epsilon D_\T{e}D_0K}{1 + \epsilon D_0K},
\end{equation}
consistent with prior results for this special case~\cite[(5.105)]{bach1969uniform}.  The factor $D_0K$ is equivalent in this case to the geometric~\cite{1956_Uzsoky_APASH} or array~\cite{lo1966optimization,pozar1984antenna} Q-factor 
\begin{equation}
    Q_\T{s} = D_0K = \frac{\sum_n |x_{n0}|^2}{\int U_0(\u{k})\,\T{d}\Omega},
    \label{eq:array-q-def}
\end{equation}
which was previously used as an indicator of dissipation and superdirectivity in studying optimal array excitations, cf.~\cite[(1)]{lo1966optimization}.

\section{Optimization Problems and Pareto-Optimal Solutions}
\label{sec:opt-problems}

We now consider the problem of finding excitation~$\M{x}$ of a given system that maximizes directivity for a predefined threshold of potential perturbation-induced reduction.  To proceed, we consider the single-parameter model for directivity reduction associated with high-directivity or ``main beam'' cases in \eqref{eq:dbar} and study the tradeoff between directivity~$D$ and sensitivity~$\sens$, which  can be written as an optimization problem
\begin{equation}
    \begin{aligned}
        	\max_{\V{J}} \quad & D \\
	\T{s.t.} \quad     & \sens = \sens_0                                                         \
    \end{aligned}
    \label{eq:opt-problem-1}
\end{equation}
Here, the degrees of freedom are represented by the current density distribution~$\V{J}$ and the parameter~$\sens_0$ sweeps along the Pareto frontier in the tradeoff space between directivity and sensitivity\footnote{Note that there exists a lower and upper bound on sensitivity~$\chi$ given by the highest and lowest eigenvalue of the generalized eigenvalue problem defined by the matrices $\M{R}$ and $\T{diag}~\M{R}$.}. This optimization problem can be written as a quadratically constrained quadratic problem (QCQP). Particularly, substituting from~\eqref{eq:Dmom} and~\eqref{eq:qsbar} and removing unnecessary constants results in
\begin{equation}
    \begin{aligned}
        	\min_{\M{x}} \quad & - \M{x}^\T{H}\M{U}^\gamma\M{x} \\
	\T{s.t.} \quad     & \M{x}^\T{H}(\T{diag}\,\M{R})\M{x} = 1                                                         \\
	                   & \M{x}^\T{H}(\T{diag}\,\M{R} - \sens_0 \M{R} )\M{x} = 0                            
    \end{aligned}
    \label{eq:opt-problem-2}
\end{equation}
where the optimized variables are expansion coefficients collected in vector~$\M{x}$. The first constraint contains a positive definite matrix, while the second constraint contains an indefinite matrix parametrized via~$\sens_0$. As formulated, this QCQP can be solved by standardized means~\cite{Nocedal2006}. In this paper, the freely downloadable solver~\cite{FunBoPackage}, described in detail in~\cite{Liska_etal_FundamentalBoundsEvaluation}, was used.

\begin{figure}
    \centering
    \includegraphics[width=\linewidth]{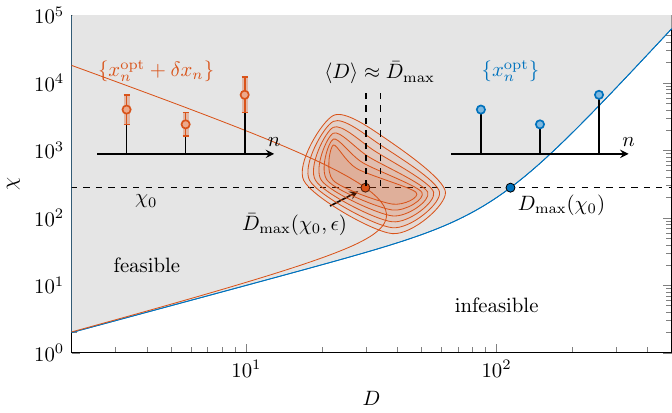}
    \caption{Schematic depiction of tradeoff space between directivity $D$ and sensitivity $
    \sens$.  The Pareto frontier (blue curve) delineates feasible and infeasible points in $D$-$\sens$ space.  Individual points (blue marker) on the Pareto frontier correspond to excitations $\M{x}_\T{opt}$ with maximum directivity $D_\T{max}(\sens_0)$ for fixed sensitivity $\sens=\sens_0$.  When perturbed by variations $\langle|\delta x_n|^2\rangle = \epsilon|x_{0n}|^2$, the anticipated directivity of the Pareto-optimal solutions is $\bar{D}_\T{max}(\sens_0,\epsilon)$ given by \eqref{eq:dbar} (red trace and marker), which approximates the ensemble average directivity $\langle D \rangle$ over many variations (red shaded region representing a two-dimensional histogram).}
    \label{fig:dq-demo}
\end{figure}

Solution of \eqref{eq:opt-problem-2} for a given value of $\sens_0$ results in an excitation $\M{x}_\T{opt}$ with sensitivity $\sens = \sens_0$ and the maximum directivity $D_\T{max}(\sens_0)$ for that particular sensitivity.  One such solution is plotted with a blue marker in Fig.~\ref{fig:dq-demo}.  Sweeping over all possible sensitivities produces the Pareto frontier drawn as a blue line in that same figure.  For a finite error $\epsilon$, the anticipated directivity~\eqref{eq:dbar} associated with any individual solution is $\bar{D}_\T{max}(\sens_0,\epsilon)$, as drawn in red.  Clearly, as directivity increases, so does the sensitivity of the excitation, eventually leading to dramatic reduction in the anticipated directivity in the presence of non-zero errors.  Subsequent sections use this format to visualize and verify trends in directivity, sensitivity, and anticipated directivity for several examples and special cases.

\subsection{Total Directivity and Realized Gain}
In cases when basis functions~$\V{\psi}_n$ in~\eqref{eq:j-basis} represent current profiles excited by transmission lines connected to specific feeding points, the directivity~$D$ might be switched to realized gain 
\begin{equation}
    G_\T{real} = 4\pi\frac{\M{x}^\T{H}\M{U}^\gamma\M{x}}{\M{x}^\T{H} \M{x}}.
    \label{eq:Greal}
\end{equation}
In such a case, the expansion coefficients~$\M{x}$ will typically be named~$\M{a}$ and would represent incident power waves in the connected transmission lines~\cite{POZAR} with incident power~\mbox{$P_\T{inc} = \M{a}^\T{H} \M{a} / 2$}.

For the case of realized gain, the optimization problem reads
\begin{equation}
    \begin{aligned}
        	\min_{\M{x}} \quad & - \M{x}^\T{H}\M{U}^\gamma\M{x} \\
	\T{s.t.} \quad     & \M{x}^\T{H} \M{x} = 1                                                         \\
	                   & \M{x}^\T{H}(\T{diag}\,\M{R} - \sens_0 \M{R} )\M{x} = 0                            
    \end{aligned}
    \label{eq:opt-problem-2real}
\end{equation}
and will generally give a different Pareto-optimal set than problem~\eqref{eq:opt-problem-2}. In the case of a dipole array studied in Section~\ref{sec:examples}, the difference between solutions to~\eqref{eq:opt-problem-2} and~\eqref{eq:opt-problem-2real} is, however, minuscule.

Similarly, in problems where total, as opposed to partial, directivity or realized gain are of interest, the preceding formulations need only be altered by considering the total radiation intensity matrix
\begin{equation}
    \M{U} = \sum_\gamma\M{U}^\gamma
\end{equation}
and its diagonal in \eqref{eq:Dmom}, \eqref{eq:zetamom}, and \eqref{eq:Greal}.  This change has minor implications for the practical solution of the optimization problems in Sec.~\ref{sec:opt-problems}, discussed in Appendix~\ref{app:direct-solution}, but no impact on the conceptual approach.

\subsection{Fast Implementation of Partial Directivity or Gain Problem}

In producing the Pareto frontier in Fig.~\ref{fig:dq-demo}, the sampling in the swept parameter $\chi_0$ is arbitrary.  This can be used advantageously to efficiently calculate the Pareto frontier by avoiding the iterations required within a general QCQP solver to hone in on a particular target value of $\chi_0$. If the entire Pareto-optimal set is desired, it suffices to evaluate $(D,\chi)$ values associated with the solutions of a parameterized system of equations
\begin{equation}
(\M{R}+\beta(\T{diag\M{R})})\M{x}_\T{opt}= \M{F}^{\gamma\T{H}},
\label{eq:direct-solution-partial}
\end{equation}
where $\beta$ is a swept auxiliary parameter. Inclusion of both polarizations results in a parameterized eigenvalue problem.  See Appendix~\ref{app:direct-solution} for further details on the derivation of \eqref{eq:direct-solution-partial} and the total directivity problem. The term $\beta(\T{diag}\M{R})$ can be interpreted as enforcing regularization.  This structure is common in many optimization problems and has appeared, in subtly different forms based on underlying assumptions, in many prior studies of tolerance in arrays, cf.~\cite[(3), case c)]{richards1975antenna} and \cite[(39)]{gilbert1955optimum}.

\section{Examples and Special Cases}
\label{sec:examples}

The following subsections contain examples demonstrating the primary features of directivity-sensitivity analysis for various radiating systems.  Codes and data to replicate selected examples are available at~\cite{dk-git}.

\subsection{Ideal Array}
We begin by considering the ``ideal array'' described in~\cite{gilbert1955optimum} with identical, independently controllable elements located at positions $\V{r}_n$ with no mutual coupling effects.  Let the basis function describing current on each element be given by a translated copy of a prototype element basis function $\V{\psi}_\T{e}(\V{r})$ nominally located at the origin, i.e.,
\begin{equation}
    \V{\psi}_n(\V{r}) = \V{\psi}_\T{e}(\V{r}-\V{r}_n)
\end{equation}
which produces the element radiation pattern $F_\T{e}^\gamma(\u{k})$.  The embedded element patterns of each element then differ only by a translation-dependent phase shift, as in \eqref{eq:element-pattern}.  Under these assumptions, the sensitivity of the ideal, uncoupled array of identical elements reads
\begin{equation}
    \sens = \frac{\int|F^\gamma_\T{e}(\u{k})|^2\,\T{d}\Omega \sum_n |x_n|^2}{\int|F^\gamma_\T{e}(\u{k})|^2|\sum x_n \T{e}^{\T{j}\V{k}\cdot\V{r}_n}|^2\,\T{d}\Omega},
\end{equation}
which has similar structure to the $K$ sensitivity factor in~\cite{gilbert1955optimum} for the same example, though here numerator and denominator are integrated over the far-field sphere and include the nominal element pattern $F_\T{e}^\gamma$.  In this case, elements of the $\M{R}$ matrix read
\begin{equation}
    R_{mn} = \frac{1}{\eta}\sum_\gamma\int|F^\gamma_\T{e}(\u{k})|^2\T{e}^{\T{j}\V{k}\cdot(\V{r}_m-\V{r}_n)}\,\T{d}\Omega
    \label{eq:rmn-ideal}
\end{equation}
For $z$-oriented Hertzian dipole elements with current $x_n$ and length $\ell$, the basis functions are
\begin{equation}
\V{\psi}_\T{e}(\V{r}) = \u{z}\ell\delta(\V{r})     
\end{equation}
with field patterns~\eqref{eq:fm} given by
\begin{equation}
    F^\theta_\T{e} = \T{j}\frac{k\eta\ell}{4\pi}\sin\theta,\quad F^\phi_\T{e} = 0
    \label{eq:f-dipole}
\end{equation}
Inserting~\eqref{eq:f-dipole} into \eqref{eq:rmn-ideal}, the elements of the matrix $\M{R}$ can be calculated as
\begin{equation}
    R_{m\neq n} = \frac{k^2\eta\ell^2}{4\pi}\left(\frac{\Delta \rho^2}{r^2}\T{j}_0(kr) + \frac{2\Delta z^2-\Delta\rho^2}{kr^3}\T{j}_1(kr)\right)
\end{equation}
and
\begin{equation}
    R_{mm} = \frac{k^2\eta\ell^2}{6\pi}
\end{equation}
with mutual distance~$r = |\V{r}_m-\V{r}_n|$, vertical displacement $\Delta z =  |\u{z}\cdot(\V{r}_m-\V{r}_n)|$, and radial displacement $\Delta \rho^2 = r^2-\Delta z^2$.

\begin{figure}
    \centering
    \includegraphics[width=\linewidth]{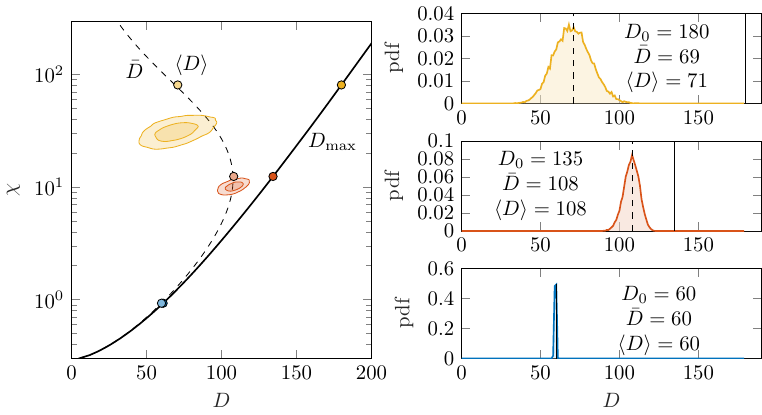}
    \caption{Directivity-sensitivity analysis of an ideal 50-element linear array of uncoupled $z$-directed Hertzian dipoles at locations $\V{r}_n = \u{x}n\lambda/5$.  Left: Pareto-optimal set of solutions (thick line) and anticipated directivity for $\epsilon = 0.02$ (thin dashed line). Three solutions (solid large markers) are selected and varied in a Monte Carlo analysis (small colored markers), with the ensemble average directivity (white markers) and 50\% and 90\% containment regions.  Right: Histograms of realized directivity from each Monte Carlo analysis along with nominal (solid line) and ensemble average (dashed line) directivities.}
    \label{fig:ideal-array-data}
\end{figure}

As an example, we consider a 50-element array of $z$-directed Hertzian dipoles located at positions $\V{r}_n = \u{x}n\lambda/10$.  To study a range of excitations for the array with varying directivity and sensitivity, the Pareto frontier of optimal excitations is calculated according to Sec.~\ref{sec:opt-problems}.  The sensitivity and directivity of this set of solutions is depicted in the left panel of Fig.~\ref{fig:ideal-array-data} as a thick line.  The anticipated directivity $\bar{D}$ is computed for each solution using \eqref{eq:dbar} and a variance parameter $\epsilon = 0.02$, drawn in Fig.~\ref{fig:ideal-array-data} as a thin dashed line.  The Pareto-optimal data exhibit a steep increase in sensitivity with respect to nominal directivity which, in the presence of even small perturbations, results in decreasing anticipated directivity for solutions of higher nominal directivity.

Three particular solutions with nominal directivity \mbox{$D_0=\left\{60,135,180\right\}$}, shown as a large colored marker on the Pareto frontier in Fig.~\ref{fig:ideal-array-data}, are selected and varied by random perturbations according to \eqref{eq:xvarprop}.  The directivity and sensitivity of each instance are recorded and represented in Fig.~\ref{fig:ideal-array-data} as shaded regions corresponding to 50\% and 90\% containment zones.  As expected, all instances sit above the Pareto frontier.  Histograms in the right panels of Fig.~\ref{fig:ideal-array-data} show the distributions of realized directivity for each selected case, along with the nominal directivity (dashed vertical line) and realized ensemble average (solid vertical line).  The ensemble average directivity for each solution is also plotted on the left panel of Fig.~\ref{fig:ideal-array-data} using the nominal sensitivity coordinate (light circles).  These data closely align with the anticipated directivity predicted by \eqref{eq:dbar}, supporting the accuracy of those assumptions.  These data also highlight the point of diminishing returns, where seeking higher nominal directivity (e.g., moving from the $D_0 = 135$ to the $D_0 = 180$ excitation) results in a lower average realized directivity under the selected error model.

\subsection{Array with Mutual Coupling}
\label{sec:fullwave-array}

The preceding example follows \cite{gilbert1955optimum} in assuming that there exists no mutual coupling between array elements.  To model a practical array in more detail, we consider a fullwave representation of a particular array including all mutual coupling effects.  

\begin{figure}
    \centering
    \includegraphics[width=\linewidth]{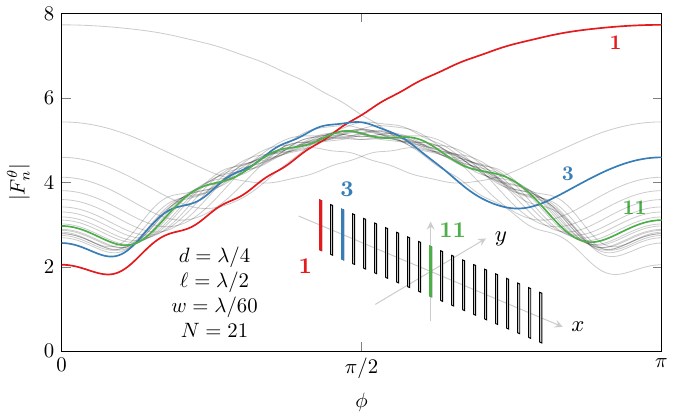}
    \caption{Embedded element patterns $F^\theta_n(\u{k})$ of a 21-element array of $z$-directed $\lambda/2$ strip dipoles at $\V{r}_n = \u{x}n\lambda/4$ with $Z_0 = 50~\Omega$ terminating impedances.  Three elements ($k = 1,3,11$) are highlighted and color-matched to their corresponding patterns.  All other element patterns are drawn in light gray.}
    \label{fig:array-eep}
\end{figure}

The array under study is a 21-element array of $z$-oriented $\lambda/2$ dipoles located at positions $\V{r}_n = \u{x}n\lambda/4$.  Each dipole is modeled as a rectangular strip with dimensions $\lambda/2\times\lambda/60$ and fed at its center by an ideal voltage source in series with a $Z_0 = 50\,\Omega$ terminating impedance.  The voltage of each driving source is taken as the controllable excitation.  As such, each embedded element pattern $F_m^\gamma(\u{k})$, shown in the $xy$-cutplane in Fig.~\ref{fig:array-eep}, is computed by applying a unit excitation to that element, while all other ports are terminated with the system impedance~$Z_0$.  The data reported in Fig.~\ref{fig:array-eep} are calculated from a method of moments (MoM) solver~\cite{atom} and were validated using a finite element method (FEM) solver~\cite{hfss2021} with average relative error of 0.5\%.  The matrix $\M{R}$ was also computed using both solvers using the procedure based on S-parameters described in Appendix~\ref{app:lossless} with an average error on the order of 3\%.  Here we reiterate that the use of embedded element patterns or S-parameters to construct the matrices $\M{U}^\gamma$ and $\M{R}$ is equivalent to using the global array current distributions associated with individual port excitations as basis functions $\V{\psi}_n$, though this intermediate step may not always be practical, depending on the selected simulation or measurement method.

\begin{figure}
    \centering
    \includegraphics[width=\linewidth]{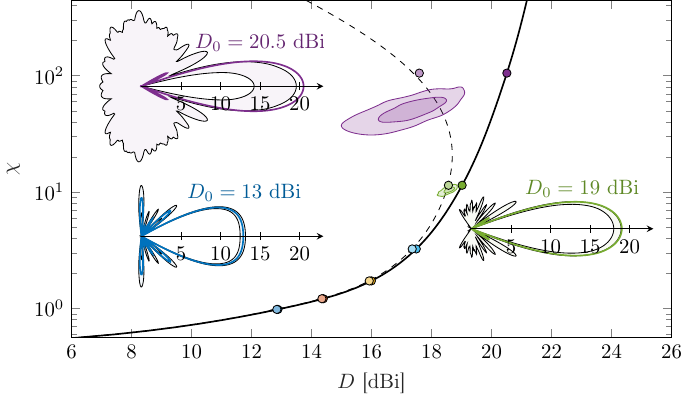}
    \caption{Directivity-sensitivity analysis of a 21-element linear array of $z$-directed $\lambda/2$ dipoles at locations $\V{r}_n = \u{x}n\lambda/4$ modeled using fullwave methods, including mutual coupling and $Z_0 = 50~\Omega$ source impedances.  Pareto-optimal set of solutions (thick black line) and anticipated directivity for $\epsilon = 0.05$ (thin black line). Six solutions (solid dark markers) are selected and varied in a Monte Carlo analysis, with the ensemble average directivity (solid light markers) and 50\% and 90\% containment regions (shaded regions).  Inset patterns for three excitations show nominal (thick lines) and ranges of realized directivity (shaded regions) from Monte Carlo analysis.}
    \label{fig:array-pareto}
\end{figure}

From these data, the Pareto-optimal set of minimum-sensitivity excitations over a range of endfire ($\phi = 0$) directivities is computed, see Fig.~\ref{fig:array-pareto}.  Following the format of Fig.~\ref{fig:ideal-array-data}, the Pareto frontier and the associated anticipated directivity for $\epsilon = 0.01$ are shown in thick and thin lines, respectively, while shaded regions show 50\% and 90\% confidence regions from Monte Carlo analyses of six particular excitations.  Insets show the nominal patterns (thick lines) and range of directivity over all Monte Carlo variations (shaded regions) for three selected excitations.

Several observations can be made from these results.  First, as in the case of an ideal array, there exists a steep cutoff in directivity after which the sensitivity sharply increases and the anticipated directivity decreases.  This is evident in the difference in distributions of directivities observed for the selected $D_0 = 19~\T{dB}$ and $D_0 = 20.5~\T{dB}$ examples, where the slight increase in directivity comes at a massive cost in sensitivity and an overall decrease in anticipated directivity under the selected $\epsilon = 0.01$ error model.  Second, the anticipated directivity $\bar{D}$ serves as a good predictor of ensemble average directivity in all six cases, though the error between these measures increases as the ensemble spread of directivities widens.  Additionally, from the inset patterns it is evident that lower-sensitivity excitations exhibit small random variations in sidelobe levels which lead to only minor variations in main-beam directivity.  The high-sensitivity pattern, however, shows much more significant sidelobe variation which ultimately leads to significant reduction in endfire directivity.

\subsection{Scaling Rules}

As in other problems leading to bounds on achievable performance, an important part of the study of $\chi$-$D$ Pareto-optimal sets in this paper is to identify scaling rules for normalizing results and predicting overall trends in key design parameters. In the context of directivity, an important scaling parameter is the electrical size~$ka$ (with~$a$ being the radius of the smallest sphere circumscribing the array), which is, for example, reflected in the so-called normal directivity~\cite{1961_Harrington}
\begin{equation}
    D_\T{normal} = \left( ka \right)^2 + 2 ka,
    \label{eq:Dnorm}
\end{equation}
which has been proposed as the limit beyond which the superdirectivity emerges and is commonly used as a benchmark for high directivity in electrically small systems~\cite{zhang2024superdirective,lynch2024super,moore2024low}.

\begin{figure}
    \centering
    \includegraphics[width=\linewidth]{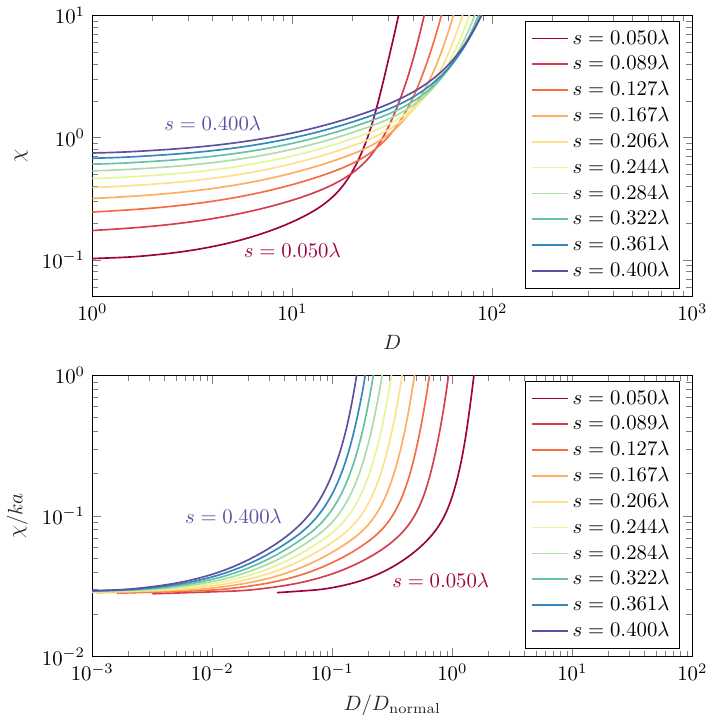}
    \caption{Directivity-sensitivity tradeoffs for a 21-element array of half-wavelength dipoles with varying spacing $s$.  Unnormalized data (top) are shown alongside the same data normalized (bottom) by Harrington's normal directivity ($D$-axis) and electrical size ($\chi$-axis).}
    \label{fig:chiD_21_kaScaling1}
\end{figure}

To see the scaling of the Pareto-optimal set studied in this paper, consider the 21-element half-wave dipole array from~Fig.~\ref{fig:array-pareto} altered with varying element spacings $s$. For each spacing value, the array will have a different overall electrical size of approximately $ka = 10.5ks\approx 66s/\lambda$. The results are depicted in~Fig.~\ref{fig:chiD_21_kaScaling1} in unnormalized (top) and normalized (bottom) formats.

The spacing of the array varies from~$s = \lambda / 20$ (smallest electrical size) to~$s = 2 \lambda / 5$ (largest electrical size). An unpleasant feature of the top panel is a strong variation of the curves with electrical size, which leads to their intersections and shifts. One possible remedy comes from the empirical observation that the lower bound of~$\chi$ approximately scales with~$ka$. Also, the directivity scaling is approximately captured in the normal directivity~\eqref{eq:Dnorm}. Normalizing the axes by these factors leads to the bottom panel of~Fig.~\ref{fig:chiD_21_kaScaling1}.

\begin{figure}
    \centering
    \includegraphics[width=\linewidth]{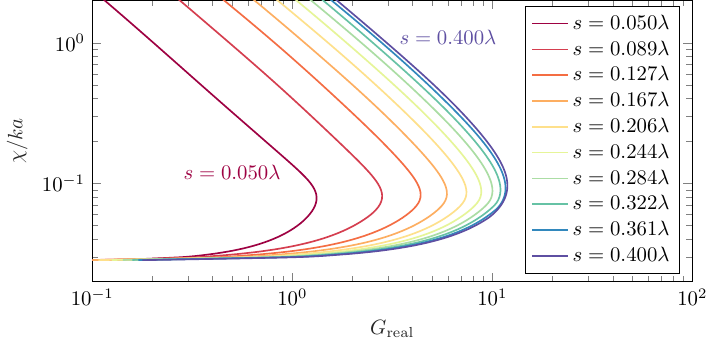}
    \caption{Realized gain and sensitivity associated with the solutions shown in Fig.~\ref{fig:chiD_21_kaScaling1}.}
    \label{fig:chiGreal_21_kaScaling}
\end{figure}

The normalized data show that larger array spacing is not favorable for the optimized trade-off. Nonetheless, this might not be an absolute statement since there are also different performance metrics of interest. For example, if realized gain is evaluated for the same Pareto-optimal set of excitations in Fig.~\ref{fig:chiD_21_kaScaling1}, we obtain the curves shown in~Fig.~\ref{fig:chiGreal_21_kaScaling}. Here, the curves also demonstrate a strong effect of dipole coupling on the matching efficiency. These curves were obtained by optimizing in the~$\chi$-$D$ space, i.e., by solving optimization problem~\eqref{eq:opt-problem-2}. Solving~\eqref{eq:opt-problem-2real} instead would be more appropriate to generate a true Pareto-optimal set in~$\chi$-$G_\T{real}$ space, but in the studied case, the difference in results is miniscule. In the presence of losses, the tradeoff between lost power and realized gain leads to qualitatively similar results.

\subsection{Comparison with Standard Array Synthesis Methods}

\begin{figure}
    \centering
    \includegraphics[width=\linewidth]{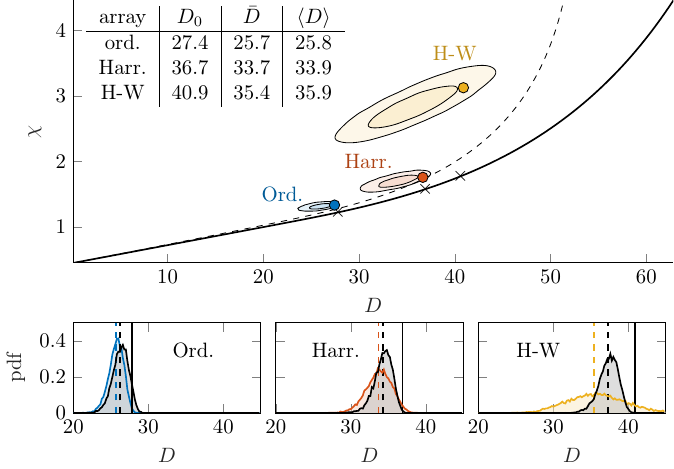}
    \caption{Directivity and sensitivity of three 21-element endfire arrays of $\lambda/2$ dipoles with $\lambda/4$ spacing: ordinary endfire (ord.), modified Harrington (Harr.), and Hansen-Woodyard (H-W).  Top panel: markers show realized array characteristics, shaded regions show 50\% and 90\% containment regions for a Monte Carlo analysis with $\epsilon = 0.05$.  $\times$ markers show Pareto optimal solutions with the same directivity for each array.  Bottom panels: histograms of Monte Carlo directivities for the realized array (color) and Pareto-optimal excitation (black). Ensemble averages (dashed vertical lines) are compared to the nominal directivity $D_0$ (solid black vertical line) in each case.}
    \label{fig:chiD_21_realizedArrays}
\end{figure}

Knowing the Pareto-optimal tradeoff between sensitivity and directivity in excitations for a particular array geometry allows for comparison with the performance of classical array designs. To that point,~Fig.~\ref{fig:chiD_21_realizedArrays} considers the same setup as in ~Fig.~\ref{fig:array-pareto} but now adds data for three realized arrays under nominal and perturbed excitation. The Hansen-Woodyard (H-W) array is defined by incident power waves are given by 
\begin{equation}
    x_n = \exp(- \T{j} \left( k s + 2.92/N \right) n),
\end{equation}
with~$s = \lambda/4$ representing array spacing. The ``Harrington'' (Harr.)  array utilizes coefficients from~\cite{1963_Harrington_TAP} which also have unit amplitude but slightly modified phase progression. Lastly, we consider an ordinary end-fire array with incident power waves given by 
\begin{equation}
    x_n = \exp(- \T{j} k s n).
\end{equation}

It can be observed that the two latter arrays sit close to the Pareto-optimal set for this array geometry.  Note that $\times$ markers show the Pareto-optimal solution corresponding to the nominal directivity of each realized array. The Hansen-Woodyard array performs worse in this respect, with sensitivity significantly higher than that of the Pareto-optimal solution of comparable directivity. The improved performance of the Pareto-optimal solution is given by non-uniform excitation amplitudes, where the phasing of the elements is very similar to that shown in~\cite{1963_Harrington_TAP}.  In all three cases, the anticipated directivity $\bar{D}$ aligns with the Monte Carlo ensemble mean directivity $\langle D \rangle$ with $\epsilon = 0.05$ to within $1\%$.

\subsection{Local Uncorrelated Variations}
\label{sec:pointwise}

The preceding examples examine tradeoffs in directivity and sensitivity from the perspective of traditional arrays, i.e., systems consisting of a collection of objects, each driven by a unique port.  While this is a useful characterization of devices with predetermined geometry, we may also be interested in the optimal $D$-$\chi$ characteristics of arbitrary current distributions within a design region $\varOmega$ before a particular geometry or port configuration has been selected, since any limits on the currents within $\varOmega$ impose physical bounds on devices realized within that region.  This strategy is widely applied in various applications in antenna design~\cite{1956_Uzsoky_APASH,1961_Harrington,capek2017minimization,GustafssonCapek2019} and other areas of applied electromagnetics~\cite{gustafsson2020upper,chao2022physical,kuang2020computational,schab2020trade}. 

One common approach to the study of optimal currents is to approximate any current within the selected region by the superposition of basis functions existing over small regions of size\footnote{Length, area, or volume; depending on the physical model being studied.} $\Delta\varOmega$, \textit{e.g.}, RWG basis functions defined over pairs of triangles~\cite{rao1982electromagnetic}.  The weighting $x_n$ of each basis function is then considered to be a completely independent degree of freedom, meaning the coefficient vector $\M{x}$ describes an approximation to a continuous current distribution within the region $\varOmega$.  Optimization problems, such as those described in Sec.~\ref{sec:opt-problems} then yield descriptions of optimal currents which in turn represent physical bounds.  

Here we study an interesting feature of $D$-$\chi$ tradeoffs of optimal currents related to the relative size $\varOmega$ of the selected basis functions. When the size of these domains are electrically small, the far-field contribution from each element reads
\begin{equation}
    x_{n0}F_n^\gamma \approx -\frac{\T{j}k\eta}{4\pi}\u{\gamma}\cdot\u{\psi}_n\T{e}^{\T{j}\phi_n}x_{n0}\Delta\varOmega
\end{equation}
and the intensity
\begin{equation}
    |x_{n0}F_n^\gamma |^2 \approx \frac{k^2\eta^2}{16\pi^2}|\u{\gamma}\cdot\u{\psi}_nx_{n0}|^2\Delta\varOmega^2.
\end{equation}
Summing contributions from all elements and taking the limit of $\Delta\varOmega\rightarrow 0$ means an integral resembling the norm of the current projected onto the polarization direction can be formed, that is,
\begin{equation}
    \lim_{\Delta\varOmega\rightarrow 0} \sum_n |x_{n0}F_n^\gamma |^2= \frac{k^2\eta^2}{16\pi^2}\Delta\varOmega\int|\u{\gamma}\cdot\V{J}|^2\,\T{d}\V{r}.
    \label{eq:small-domain-scaling}
\end{equation}
However, this term vanishes in the limit of increasingly small basis domains.  Integrating over the far-field sphere doesn't affect the overall scaling in domain size $\Delta\varOmega$, indicating that both the numerator of \eqref{eq:qsbar} and the diagonal elements $R_{mm}$ scale as $\Delta\varOmega$.  In contrast, the denominator of \eqref{eq:qsbar} represents a total radiated power which, for sufficiently small domain size, should be invariant in that parameter for a fixed current density~$\V{J}$.  Therefore, we expect the sensitivity $\sens$ to scale linearly with the domain size $\Delta\varOmega$.

Physically, this can be understood as the individual contributions to the background intensity becoming smaller and smaller, eventually smoothing out to zero for pointwise variations where neighboring errors, on average, cancel each other out in the far field.  Alternatively, if the basis size $\Delta\varOmega$ is taken as a relative measure of the correlation length, area, or volume in the adopted error model, then it follows that the sensitivity of the system is directly proportional to this dimension.  This indicates that it is not the variation $\delta\V{J}$ alone which impacts directivity sensitivity, but the correlation length scale of that error, as described by the characteristic length $\Delta\varOmega$.

\begin{figure}
    \centering
    \includegraphics[width=\linewidth]{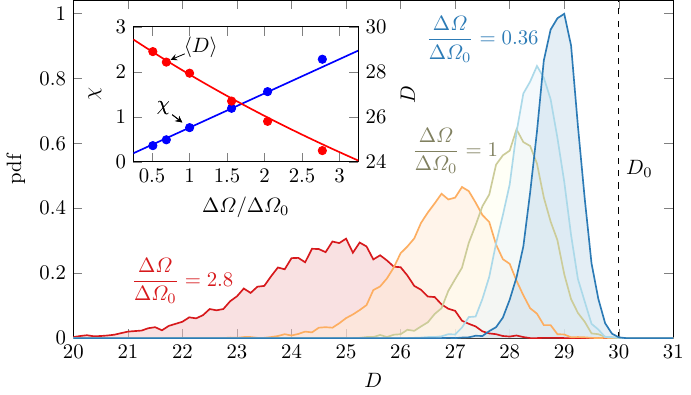}
    \caption{Directivity histograms for a current density confined to a 2:1 rectangular region of electrical size $ka = 5$ with nominal directivity $D_0 = 30$ subject to random variations with $\epsilon = 0.1$.  Various mesh densities are described by relative triangle area $\Delta \varOmega / \Delta \varOmega_0$.  Anticipated directivity and sensitivity for this current distribution at varying mesh densities are shown in the inset.  Markers represent numerically calculated sensitivity and ensemble average directivity, while lines represent trends predicted by \eqref{eq:small-domain-scaling} and \eqref{eq:dbar}.}
    \label{fig:rwg-pareto}
\end{figure}

To examine this numerically, we consider a rectangular domain with $2:1$ aspect ratio and electrical size $ka = 5$ represented by varying densities of RWG basis functions~\cite{rao1982electromagnetic}, each characterized by uniform triangle area $\Delta\varOmega$ relative to a central value of $\Delta\varOmega_0 = 0.0025\lambda^2$.  Directivity is measured in the endfire direction relative to the long dimension of the rectangular domain.  In each basis, currents lying on the Pareto boundary in directivity-sensitivity space are computed using the methods described in Sec.~\ref{sec:opt-problems}.  Solutions for nominal directivity $D_0 = 30$ are selected and subject to Monte Carlo variations with normally distributed errors characterized by $\epsilon = 0.1$.  The probability densities of directivity are shown in Fig.~\ref{fig:rwg-pareto}, where it is evident that decreasing domain size leads to lower variation in directivity.  Also shown are the ensemble average directivity and sensitivity for each mesh density, which align well with the scaling predicted in \eqref{eq:small-domain-scaling} and the associated anticipated directivity $\bar{D}$.

\subsection{Effect of Error Distributions}

The special case described in Sec.~\ref{sec:zero-mean} and used in all preceding example calculations made three assumptions on the nature of errors in the realized current distribution.  First, it was assumed that the individual error terms applied to each basis function are uncorrelated.  Second, these error terms were assumed to be zero-mean.  Finally, the variance of each error term is defined in terms of the expansion coefficient, i.e., \eqref{eq:xvarprop}, corresponding to a fractional change from its nominal value.  While these assumptions may not be applicable to all physical systems, their explicit use is nevertheless common in prior work on array sensitivity~\cite{gilbert1955optimum}, and implicitly used wherever array Q-factor (as defined in \cite{1956_Uzsoky_APASH}) is used as a measure of array tolerance, e.g., \cite{lo1966optimization}.

Notably these assumptions do not prescribe particular error distribution (e.g., Gaussian, uniform), but specify only mean and variance statistics.  This implies that results regarding the anticipated directivity~\eqref{eq:dbar} are independent of distribution type when these parameters are held constant.  

\begin{figure}
    \centering
    \includegraphics[width=\linewidth]{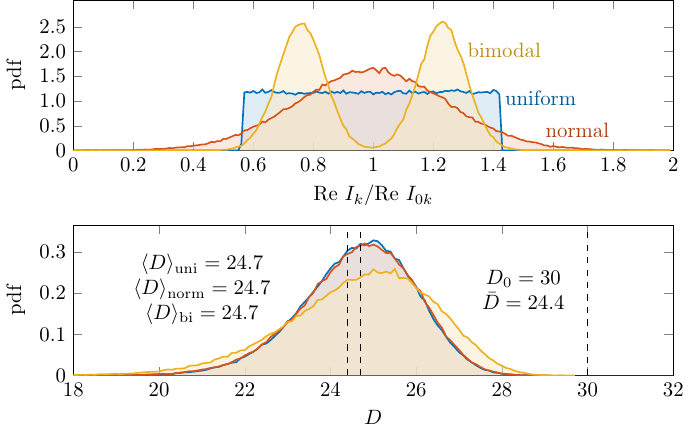}
    \caption{(Top) Histogram depicting variation in the real part of one expansion coefficient $x_k$ subject to zero-mean errors with uniform, normal, and bimodal distributions.  All three distributions have the same variance set by \eqref{eq:xvarprop} and $\epsilon = 0.1$. (Bottom) Histogram depicting directivity distributions under the error models depicted above.  The nominal directivity for this example is $D_0 = 30$. The anticipated directivity and ensemble average directivity of all three error models are also listed.}
    \label{fig:distributions}
\end{figure}

To numerically verify this result, consider the case of optimal currents on the rectangular design region in the preceding example.  The current producing nominal directivity $D_0 = 30$ on the coarse region size $\Delta\varOmega / \Delta\varOmega_0 = 2.8$ is again subject to random variations defined by the variance scaling $\epsilon = 0.1$.  In this study, however, three distributions, all sharing the same zero-mean and variance are considered.  Histograms of the real part of one particular expansion coefficient $x_k$ are shown in Fig.~\ref{fig:distributions}, showing the three cases of uniform, normal, and bimodal error distributions over 200,000 random trials.  The corresponding directivity histograms are shown in the bottom panel of that same figure.  Despite the vastly different error distributions, all three models result in similar ensemble-mean directivity $\langle D\rangle$, which itself closely aligns with the anticipated directivity $\bar{D}$ predicted by \eqref{eq:dbar}, which depends only on each distribution's variance and zero-mean property.

\subsection{Optimal Currents in Orthogonal Far Field Basis}
Due to the diagonal operator present in the sensitivity and uniformity parameters, those metrics (and the resulting anticipated directivity) are necessarily basis-dependent.  Certain choices of basis functions can lead to notable properties, particularly in regard to the sensitivity metric.  Consider current profiles~$\V{\psi}_n$ in~\eqref{eq:j-basis} with the orthogonality property
\begin{equation}
    \int \V{F}_m^*\cdot\V{F}_n\,\T{d}\Omega = \delta_{mn}
    \label{eq:ff-orth}
\end{equation}
where $\delta_{mn}$ is the Kronecker delta function. The total far field is expressed as a sum
\begin{equation}
    \V{F}_0 = \sum x_n \V{F}_n
\end{equation}
and, due to the orthogonality property in \eqref{eq:ff-orth}, the integrated total field reduces to a sum of squared coefficient magnitudes, i.e.,
\begin{equation}
    \int |\V{F}_0|^2 \,\T{d}\Omega = \sum |x_n|^2,
\end{equation}
If these expansion coefficients are subject to variation of the form of \eqref{eq:err-1} and \eqref{eq:xvarprop}, the above diagonalization of the radiated-power matrix~$\M{R}$ reduces the sensitivity term to
\begin{equation}
    \sens = 1,
    \label{eq:dk-1}
\end{equation}
regardless of directivity.

\begin{figure}
    \centering
    \includegraphics[width=\linewidth]{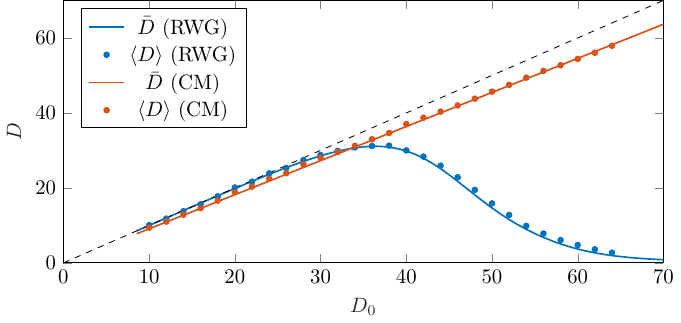}
    \caption{Anticipated directivity (solid lines) and ensemble average directivity (markers) for current distributions expressed in RWG (828 basis functions, blue) and characteristic mode (143 modes, red) bases with variance parameter $\epsilon = 0.1$.  Ideal directivity with no variance (dashed black line) is shown for reference.  Each ensemble average is generated by Monte Carlo analysis.}
    \label{fig:cm-basis}
\end{figure}

As an example, consider the rectangular region studied in Sec.~\ref{sec:pointwise}.  Using the  middle of the three mesh densities shown in Fig.~\ref{fig:rwg-pareto}, the system is represented by 828 RWG basis functions.  From the corresponding impedance matrix, the characteristic modes of the system are computed, of which the first 143 have eigenvalues within a dynamic range limited by numerical precision and strictly obey the orthogonality property in \eqref{eq:ff-orth}.  According to \eqref{eq:dk-1}, the anticipated directivity $\bar{D}$ of any current distribution with nominal directivity $D$ should equal
\begin{equation}
    \bar{D} = \frac{D_0}{1+\epsilon}
    \label{eq:dbar-cm}
\end{equation}
when the variation $\epsilon$ is applied to the modal weights used to expand the current in characteristic modes.  To study this, we consider the set of Pareto-optimal currents in Fig.~\ref{fig:rwg-pareto} as a set of arbitrary currents with varying directivity.  The ensemble average directivity of each solution is computed using a Monte Carlo analysis with variance parameter $\epsilon = 0.1$ for the currents expressed in the original RWG basis and in the characteristic mode basis.  Results are shown in Fig.~\ref{fig:cm-basis}, where ensemble average directivities are shown (solid markers) overlaid on the anticipated directivity (thin lines) computed in both bases.  While not rigorously equal, the results show that the anticipated directivity $\bar{D}$ is a good predictor of ensemble average directivity for both bases.  Further, the anticipated directivity for a far-field orthogonal basis \eqref{eq:dbar-cm} exhibits less penalization in the high-directivity limit, relative to the local RWG basis.

\subsection{Accuracy of Anticipated Directivity Estimates}
\label{sec:estimate-accuracy}

Throughout all preceding examples, we primarily studied the average reduction in main-beam directivity using the approximation in \eqref{eq:dbar}, which depends only on the directivity $D_0$ and sensitivity $\chi$ and lends itself well to the study of tradeoffs between these parameters.  While this approximation involves an assumption of relatively high directivity in the observation direction (relative to error terms), all numerical examples demonstrated reasonable agreement between that estimate and ensemble average results obtained from Monte Carlo analyses.  

\begin{figure}
    \centering
    \includegraphics[width=\linewidth]{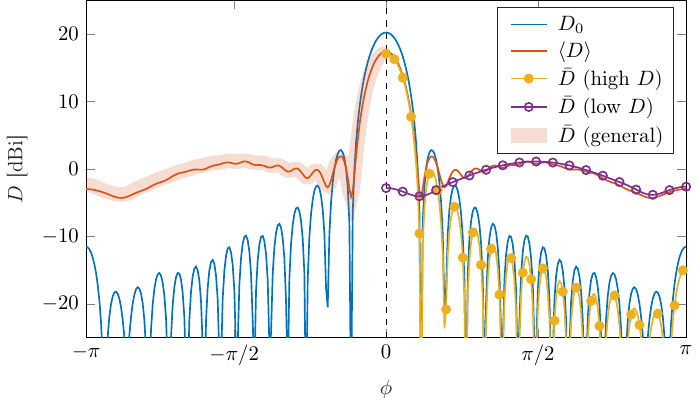}
    \caption{Comparison of estimate accuracy between anticipated directivity $\bar{D}$ and the ensemble mean directivity $\langle D \rangle$ using the general \eqref{eq:dbar-complete} [left], high-directivity \eqref{eq:dbar} [right], and the low-directivity \eqref{eq:dbar-null} [right] estimates.  Data are based on the $D_0 = 20.5~\T{dB}$ example in Fig.~\ref{fig:array-pareto}.}
    \label{fig:estimate-accuracy}
\end{figure}

To more thoroughly examine the validity of this approximation, where it breaks down, and how it can be supplemented with the rigorous model in \eqref{eq:dbar-complete} or the ``null'' approximation in \eqref{eq:dbar-null}, we again consider the 21-element array in Sec.~\ref{sec:fullwave-array} and the specific excitation associated with a nominal endfire directivity $D_0 = 20.5~\T{dB}$.   As in Fig.~\ref{fig:array-pareto}, the nominal excitation is then randomly varied using a normal distribution with variance set by the parameter $\epsilon =0.01$.  

The nominal directivity pattern $D_0$ for this excitation is plotted (blue) along with the ensemble mean directivity $\langle D \rangle$ is recorded at each angle (red) in Fig.~\ref{fig:estimate-accuracy}.  On the left side of that figure, the general three-parameter approximation of anticipated directivity $\bar{D}$ is shown to have excellent agreement with the ensemble average directivity $\langle D \rangle$ across all observation angles, regardless if the deviation from the nominal pattern is a decrease (near the main beam at $\phi = 0$) or an increase of side- and backlobes (away from $\phi = 0$).  In contrast, the right side of Fig.~\ref{fig:estimate-accuracy} shows the same nominal and ensemble average directivity patterns against the low- and high-directivity estimates \eqref{eq:dbar-null} and \eqref{eq:dbar}, respectively.  As expected, the two-parameter model for decreases in high-directivity directions agrees well near the main beam but breaks down in the side- and backlobe regions.  The opposite holds for the low-directivity estimate.  

While the general three-parameter model provides the most robust estimate of deviations in directivity, the two simpler, more specific estimates may be of interest in forming optimization problems studying tradeoffs in main-beam and sidelobe sensitivities.  Analyses and examples such as these are beyond the scope of this work, but may be of interest in future studies.

\section{Conclusion}
In this work, we have derived predictors of mean directivity of radiators subject to random errors.  Here we primarily consider errors in a system's excitation (e.g., element excitation amplitude or modal weighting coefficient), though in special cases these errors are equivalent to specific forms of mechanical error (e.g., displacement).  The formulation here generalizes prior work to account for heterogeneous arrays, continuous currents described by modal expansions, and systems involving correlated errors.  The uniformity and sensitivity measures introduced here are expressed in terms of readily-available data, making them applicable to simulated models or measurements of realized systems.  The optimization of these metrics and the study of their tradeoffs is straightforward to implement, allowing for fast quantitative tolerance analysis without the need for Monte Carlo simulations.

The single mandatory approximation made in \eqref{eq:anticipD1} is supported by the close agreement between Monte Carlo ensemble mean directivities and the anticipated directivity $\bar{D}$ in all examples studied here.  Similarly, the high-directivity approximation~\eqref{eq:dbar}, which involves only nominal directivity $D_0$ and sensitivity $\chi$ and is particularly well-suited to multi-objective optimization, is also shown to be accurate in cases involving the analysis of main-beam characteristics of high-directivity systems

The zero-mean uncorrelated error model described in Sec.~\ref{sec:zero-mean} is not mandatory, but it greatly simplifies the general uniformity and sensitivity measures for correlated arbitrary-mean errors.  Future work could study the applicability of both classes of models to practical systems involving manufacturing, excitation, and measurement errors.  The conclusions of any such study would likely be system- and application-specific.  Additionally, further work could study the statistical properties of realized directivities beyond mean values, e.g., variance, in order to provide yield estimates in randomly varying ensembles of devices.

\appendices

\section{Special Case of Lossless Antenna Arrays}
\label{app:lossless}
All prior discussion is general for any current basis of the form of \eqref{eq:j-basis}, however some special relations and simplifications are possible for the case of a lossless antenna array driven at a set of discrete ports.  In this case, directivity and sensitivity can be directly computed from port quantities and simple far-field data, allowing for the construction of the $\M{R}$ and $\M{U}^\gamma$ matrices using arbitrary simulation tools or measured data.  

Let an $N$-element array be interfaced with transmission lines of characteristic impedance $Z_0$.  Denote the incoming power wave coefficients on each port by the coefficients $\M{a}$, along with the corresponding outgoing wave coefficients $\M{b} = \M{S}\M{a}$, where the $\M{S}$ contains the system's S-parameters.  Here we adopt the normalization that incident power at port $n$ is given by $|a_n|^2/2$. 
For lossless systems, conservation of power requires
\begin{equation}
    P_\T{rad} = \frac{1}{2} \M{a}^\T{H}(\M{1-\M{S}^\T{H}\M{S}})\M{a}
    \label{eq:lossless-power-cons}
\end{equation}
so that the matrix $\M{R}$ in the selected basis reads
\begin{equation}
    \M{R} = \M{1}-\M{S}^\T{H}\M{S}
    \label{eq:lossless-r-from-s}
\end{equation}

The quantity $F_n^\gamma$ is the far field of the system measured in the observation direction when port $n$ is excited with unit incident power wave, i.e., $a_m = \delta_{mn}$.  This is equivalent to the embedded element pattern of port $n$.  For arbitrary excitations, the form of \eqref{eq:fm} and \eqref{eq:u-op} applies directly and the matrix $\M{U}^\gamma$ in this basis is given by \eqref{eq:u-op} where again the row vector $\M{F}^\gamma$ contains entries $F^\gamma_n$ corresponding to the embedded element patterns of each element.

The relations in \eqref{eq:lossless-power-cons} and \eqref{eq:lossless-r-from-s} hold rigorously only for lossless systems.  However, it may serve as a fair approximation of low-loss devices.  The significance of relation \eqref{eq:lossless-r-from-s} features prominently elsewhere in the calculation of other field quantities, e.g., envelope correlation coefficients~\cite{stein1962cross,salonen2002estimation,thaysen2006envelope}, using port measurements in place of costly or impractical measurements of radiated field data over the entire far-field sphere, either in simulation or measurement.  Hence, for lossless or low-loss systems, the construction of the matrix~$\M{R}$ via \eqref{eq:lossless-r-from-s} provides a convenient shortcut, which, together with \eqref{eq:u-op} allows for the calculation of directivity and sensitivity of arbitrary excitations using only $N$-port S-parameter data and measurement of $N$ embedded element patterns at a single observation angle.


\section{Parametric Sweep of Pareto-Optimal Solutions}
\label{app:direct-solution}
The problem in \eqref{eq:opt-problem-2} and \eqref{eq:opt-problem-2real} can be solved for a particular value of sensitivity $\chi_0$ using a general-purpose QCQP solver. For each value of the sensitivity, the process reduces to an iteratively-solved optimization problem in a single Lagrange parameter. The iteration is needed to find the point corresponding to the target sensitivity $\chi_0$. However, in many cases, the entire Pareto frontier is of interest for examining trade-offs between directivity and sensitivity.  In such a case, the repeated iterative procedures can be replaced by a single parameterized sweep over an auxiliary variable to produce the entire Pareto-optimal set in an efficient, direct manner.

This strategy is clear when examining the Lagrangian of the optimization problem in \eqref{eq:opt-problem-2}
\begin{equation}
    \mathcal{L}(\mu,\nu,\M{x}) = - \M{x}^\T{H}(\M{U}+\mu\M{R}+\nu(\T{diag} \,\M{R}))\M{x} + \dfrac{1}{\chi_0}\mu + \nu,
\end{equation}
where $\mu$ and $\nu$ are Lagrange multipliers, see~\cite{boyd2004convex}.  The stationary points of the Lagrangian are the solutions to 
\begin{equation}
    \M{U} \M{x}_\T{opt} = \alpha \left( \M{R} + \beta (\T{diag} \,\M{R}) \right) \M{x}_\T{opt},
    \label{eq:app-stationary}
\end{equation}
which is an eigenvalue problem with eigenvalue $\alpha = -\mu$, which is parametrized by the parameter $\beta = \nu/\mu$. The sensitivity $\chi_0$ does not appear in the above expression, and the parameter $\beta$ must be swept to obtain a stationary point maximizing the term  $ - \alpha \left( {1/{\chi _0} + \beta } \right)$ under the condition that the matrix $\M{U} - \alpha \left( \M{R} + \beta\,\T{diag} \,\M{R} \right)$ is negative definite.  

If the specific value of sensitivity $\chi_0$ is not desired, the eigenvalue problem~\eqref{eq:app-stationary} can just be swept in the full allowed range of parameter~$\beta$. The two non-zero (note that matrix~$\M{U}$ is at most rank-2) solutions to~\eqref{eq:app-stationary} generate specific pairs of values~$D-\chi$ from which the Pareto-optimal set can be constructed. For problems involving partial directivity, the matrix $\M{U}$ is of rank-1, and the eigenvector corresponding to the sole non-zero eigenvalue of~\eqref{eq:app-stationary} is the solution to the parameterized system of equations \eqref{eq:direct-solution-partial}.

A similar technique was employed to study the tradeoffs between directivity and sensitivity using the simplified sensitivity operator~\eqref{eq:k-def}~\cite{gilbert1955optimum,cox1986practical} or the corresponding array Q-factor~\eqref{eq:array-q-def}~\cite{lo1966optimization}.

\bibliographystyle{ieeetr}
\bibliography{ref_gain,extraRefs, capekRef}

@Article{GustafssonCapek2019,
  author  = {Gustafsson, M. and Capek, M.},
  title   = {Maximum Gain, Effective Area, and Directivity},
  journal = {IEEE Trans. Antennas Propag.},
  year    = {2019},
  volume  = {67},
  number  = {8},
  month   = {Aug.},
  pages   = {5282 -- 5293},
  doi     = {10.1109/TAP.2019.2916760},
}

@ARTICLE{HarringtonMautz1971a,
  author = {Harrington, R. F. and Mautz, J. R.},
  title = {Theory of Characteristic Modes for Conducting Bodies},
  journal = {IEEE Trans. Antennas Propag.},
  year = {1971},
  volume = {19},
  pages = {622--628},
  number = {5},
  month = {Sept.},
}

@Article{Liska_etal_FundamentalBoundsEvaluation,
  author  = {Jakub Liska and Lukas Jelinek and Miloslav Capek},
  title   = {Fundamental Bounds to Time-Harmonic Quadratic Metrics in Electromagnetism: {O}verview and Implementation},
  note = {eprint arXiv: 2110.05312},
  doi     = {10.48550/arXiv.2110.05312},
  journal = {arXiv},
  year    = {2021},
  url     = {https://arxiv.org/abs/2110.05312},
}

@article{FunBoPackage,
  author = {Jakub Liska and Lukas Jelinek and Miloslav Capek},
  title  = {Fundamental Bounds Package},
  date   = {17.01.2026},
  note    = {http://www.antennatoolbox.com/fundamentalBounds},
}

@BOOK{Nocedal2006,
  title     = "Numerical Optimization",
  author    = "Nocedal, Jorge and Wright, Stephen",
  publisher = "Springer",
  series    = "Springer Series in Operations Research and Financial Engineering",
  edition   =  2,
  month     =  jan,
  year      =  2006,
  language  = "en"
}

@article{krasnok2026advanced,
  title={Advanced Superdirective Antennas},
  author={Krasnok, Alex},
  journal={arXiv preprint arXiv:2602.04121},
  year={2026}
}

@article{yaghjian2008electrically,
  title={Electrically small supergain end-fire arrays},
  author={Yaghjian, Arthur D and O'Donnell, Terry H and Altshuler, Edward E and Best, Steven R},
  journal={Radio Science},
  volume={43},
  number={03},
  pages={1--13},
  year={2008},
  publisher={AGU}
}

@ARTICLE{elliott1958mechanical,
  author={Elliott, R.},
  journal={IRE Transactions on Antennas and Propagation}, 
  title={Mechanical and electrical tolerances for two-dimensional scanning antenna arrays}, 
  year={1958},
  volume={6},
  number={1},
  pages={114-120},
  doi={10.1109/TAP.1958.1144549}}

@ARTICLE{pozar1984antenna,
  author={Pozar, D.},
  journal={IEEE Transactions on Antennas and Propagation}, 
  title={Antenna synthesis and optimization using weighted Inagaki modes}, 
  year={1984},
  volume={32},
  number={2},
  pages={159-165},
  doi={10.1109/TAP.1984.1143292}}

@article{william2025comparative,
  title={A Comparative Review on Superdirectivite Antenna Systems in Modern Literature},
  author={William, Monica Wasfy and Antar, Yahia MM and Freundorfer, Alois P},
  journal={IEEE Access},
  year={2025},
  publisher={IEEE}
}

@BOOK{1961_Harrington,
  title     = "Time-harmonic electromagnetic fields",
  author    = "Harrington, Roger Fuller",
  publisher = "McGraw-Hill",
  series    = "Electrical \& Electronic Engineering S.",
    year      =  1961,
  address   = "New York, NY"
}

@article{1963_Harrington_TAP,
  title = {Modified progressive phasing for endfire arrays},
  volume = {11},
  ISSN = {0096-1973},
    DOI = {10.1109/tap.1963.1137970},
  number = {1},
  journal = {IEEE Transactions on Antennas and Propagation},
  publisher = {Institute of Electrical and Electronics Engineers (IEEE)},
  author = {Harrington,  R.},
  year = {1963},
  month = jan,
  pages = {13–17}
}

@article{bach1969uniform,
  title={Uniformly spaced arrays},
  author={Bach, H. and Hansen J.E.},
  journal={Antenna Theory},
  volume={1},
  pages={139--206},
  year={1969},
  publisher={McGraw-Hill New York}
}

@article{skolnik1969nonuniform,
  title={Uniformly spaced arrays},
  author={Skolnik, M.},
  journal={Antenna Theory},
  volume={1},
  pages={207--234},
  year={1969},
  publisher={McGraw-Hill New York}
}

@article{lee2021accurate,
  title={Accurate statistical model of radiation patterns in analog beamforming including random error, quantization error, and mutual coupling},
  author={Lee, Seunghoon and Song, Ho-Jin},
  journal={IEEE Transactions on Antennas and Propagation},
  volume={69},
  number={7},
  pages={3886--3898},
  year={2021},
  publisher={IEEE}
}

@article{sjoberg2005coherent,
  title={Coherent effects in single scattering and random errors in antenna technology},
  author={Sj\"{o}berg, Daniel},
  journal={Progress In Electromagnetics Research},
  volume={50},
  pages={13--39},
  year={2005},
  publisher={EMW Publishing}
}

@article{schab2020trade,
  title={Trade-offs in absorption and scattering by nanophotonic structures},
  author={Schab, Kurt and Rothschild, Austin and Nguyen, Kristi and Capek, Miloslav and Jelinek, Lukas and Gustafsson, Mats},
  journal={Optics Express},
  volume={28},
  number={24},
  pages={36584--36599},
  year={2020},
  publisher={Optical Society of America}
}

@article{richards1975antenna,
  title={Antenna pattern synthesis based on optimization in a probabilistic sense},
  author={Richards, W and Lo, Yuen},
  journal={IEEE Transactions on Antennas and Propagation},
  volume={23},
  number={2},
  pages={165--172},
  year={1975},
  publisher={IEEE}
}

@ARTICLE{lee2005decision,
  author={Jeongheum Lee and Yongbeum Lee and Hyeongdong Kim},
  journal={IEEE Transactions on Antennas and Propagation}, 
  title={Decision of error tolerance in array element by the Monte Carlo method}, 
  year={2005},
  volume={53},
  number={4},
  pages={1325-1331},
  doi={10.1109/TAP.2005.844444}}

@ARTICLE{anselmi2013tolerance,
  author={Anselmi, Nicola and Manica, Luca and Rocca, Paolo and Massa, Andrea},
  journal={IEEE Transactions on Antennas and Propagation}, 
  title={Tolerance Analysis of Antenna Arrays Through Interval Arithmetic}, 
  year={2013},
  volume={61},
  number={11},
  pages={5496-5507},
  doi={10.1109/TAP.2013.2276927}}

@INPROCEEDINGS{moghaddam2019statistical,
  author={Moghaddam, Mohammad Hossein and Rezaei Aghdam, Sina and Eriksson, Thomas},
  booktitle={2019 IEEE Global Conference on Signal and Information Processing (GlobalSIP)}, 
  title={Statistical Analysis of Antenna Array Systems with Perturbations in Phase, Gain and Element Positions}, 
  year={2019},
  volume={},
  number={},
  pages={1-5},
  doi={10.1109/GlobalSIP45357.2019.8969083}}

@article{zhang2024superdirective,
  title={Superdirective, electrically small, endfire-radiating Huygens quadrupole antenna},
  author={Zhang, Zhehao and Li, Mei and Dai, Qi and Ziolkowski, Richard W},
  journal={IEEE Transactions on Antennas and Propagation},
  volume={72},
  number={10},
  pages={7615--7627},
  year={2024},
  publisher={IEEE}
}

@article{cox1986practical,
  title={Practical supergain},
  author={Cox, Henry and Zeskind, R and Kooij, Theo},
  journal={IEEE Transactions on Acoustics, Speech, and Signal Processing},
  volume={34},
  number={3},
  pages={393--398},
  year={1986},
  publisher={IEEE}
}

@article{diao2017practical,
  title={Practical superdirectivity with resonant screened apertures motivated by a poynting streamlines analysis},
  author={Diao, Junming and Warnick, Karl F},
  journal={IEEE Transactions on Antennas and Propagation},
  volume={66},
  number={1},
  pages={432--437},
  year={2017},
  publisher={IEEE}
}

@article{hsiao1985design,
  title={Design of error tolerance of a phased array},
  author={Hsiao, JK},
  journal={Electronics Letters},
  volume={21},
  number={19},
  pages={834--836},
  year={1985},
  publisher={IET}
}

@article{howard1967statistical,
  title={Statistical patterns of a general array},
  author={Howard, J},
  journal={IEEE Transactions on Antennas and Propagation},
  volume={15},
  number={1},
  pages={60--65},
  year={1967},
  publisher={IEEE}
}

@inproceedings{rondinelli1966effects,
  title={Effects of random errors on the performance of antenna arrays of many elements},
  author={Rondinelli, L},
  booktitle={1958 IRE International Convention Record},
  volume={7},
  pages={174--189},
  year={1966},
  organization={IEEE}
}

@article{trucco2022statistics,
  title={The statistics of superdirective beam patterns},
  author={Trucco, Andrea},
  journal={IEEE Transactions on Signal Processing},
  volume={70},
  pages={1959--1975},
  year={2022},
  publisher={IEEE}
}

@inproceedings{moore2024low,
  title={Low-profile super-realised gain antennas},
  author={Moore, James and Graham, Aaron M and Tentzeris, Manos M and Fusco, Vincent and Asimonis, Stylianos D},
  booktitle={2024 18th European Conference on Antennas and Propagation (EuCAP)},
  pages={1--4},
  year={2024},
  organization={IEEE}
}

@inproceedings{lynch2024super,
  title={Super-realized gain {Huygens} antennas},
  author={Lynch, Donal Patrick and Fusco, Vincent and Tentzeris, Manos M and Asimonis, Stylianos D},
  booktitle={2024 18th European Conference on Antennas and Propagation (EuCAP)},
  pages={1--3},
  year={2024},
  organization={IEEE}
}

@phdthesis{ruze1952physical,
  title={Physical limitations on antennas},
  author={Ruze, John},
  year={1952},
  school={Massachusetts Institute of Technology}
}

@article{gustafsson2020upper,
  title={Upper bounds on absorption and scattering},
  author={Gustafsson, Mats and Schab, Kurt and Jelinek, Lukas and Capek, Miloslav},
  journal={New Journal of Physics},
  volume={22},
  number={7},
  pages={073013},
  year={2020},
  publisher={IOP Publishing}
}

@book{boyd2004convex,
  title={Convex optimization},
  author={Boyd, Stephen and Vandenberghe, Lieven},
  year={2004},
  publisher={Cambridge university press}
}

@article{chao2022physical,
  title={Physical limits in electromagnetism},
  author={Chao, Pengning and Strekha, Benjamin and Kuate Defo, Rodrick and Molesky, Sean and Rodriguez, Alejandro W},
  journal={Nature Reviews Physics},
  volume={4},
  number={8},
  pages={543--559},
  year={2022},
  publisher={Nature Publishing Group UK London}
}

@article{capek2017minimization,
  title={Minimization of antenna quality factor},
  author={Capek, Miloslav and Gustafsson, Mats and Schab, Kurt},
  journal={IEEE Transactions on Antennas and Propagation},
  volume={65},
  number={8},
  pages={4115--4123},
  year={2017},
  publisher={IEEE}
}

@article{kuang2020computational,
  title={Computational bounds to light--matter interactions via local conservation laws},
  author={Kuang, Zeyu and Miller, Owen D},
  journal={Physical Review Letters},
  volume={125},
  number={26},
  pages={263607},
  year={2020},
  publisher={APS}
}

@article{cheng2005optimization,
  title={Optimization techniques for antenna arrays},
  author={Cheng, David K},
  journal={Proceedings of the IEEE},
  volume={59},
  number={12},
  pages={1664--1674},
  year={1971},
  publisher={IEEE}
}

@Article{LebedevLaikov_QuadratureRuleSphere,
  author       = {Lebedev, V. I. and Laikov, D. N.},
  date         = {1999},
  journaltitle = {Doklady Mathematics},
  title        = {A Quadrature Formula for the Sphere of the 131st Algebraic Order of Accuracy},
  number       = {3},
  pages        = {477--481},
  volume       = {59},
}

@book{gray2004introduction,
  title={An introduction to statistical signal processing},
  author={Gray, Robert M and Davisson, Lee D},
  year={2004},
  publisher={Cambridge University Press}
}

@article{rao1982electromagnetic,
  title={Electromagnetic scattering by surfaces of arbitrary shape},
  author={Rao, Sadasiva and Wilton, Donald and Glisson, Allen},
  journal={IEEE Transactions on antennas and propagation},
  volume={30},
  number={3},
  pages={409--418},
  year={1982},
  publisher={IEEE}
}

@Electronic{dk-git,
  title        = {Directivity and Error Sensitivity in Arrays and Continuous Sources},
  note          = {https://github.com/kschab/directivity-sensitivity},
}

@Electronic{atom,
  title        = {{A}ntenna {T}oolbox for {MATLAB} ({AToM})},
  year         = {2022},
  url          = {{www.antennatoolbox.com}},
  note         = {{www.antennatoolbox.com}},
  organization = {Czech Technical University in Prague},
}

@Electronic{hfss2021,
  title        = {{HFSS}},
  url          = {www.ansys.com},
  organization = {Ansys},
  year         = {2022},  
  version = {2021 R1}
}

@inproceedings{salonen2002estimation,
  title={Estimation of signal correlation in antenna arrays},
  author={Salonen, Ilkka and Vainikainen, Pertti},
  booktitle={Proc. JINA},
  volume={2},
  pages={383--386},
  year={2002}
}

@ARTICLE{stein1962cross,
  author={Stein, S.},
  journal={IRE Transactions on Antennas and Propagation}, 
  title={On cross coupling in multiple-beam antennas}, 
  year={1962},
  volume={10},
  number={5},
  pages={548-557},
  doi={10.1109/TAP.1962.1137917}}

@article{thaysen2006envelope,
  title={Envelope correlation in (N, N) MIMO antenna array from scattering parameters},
  author={Thaysen, Jesper and Jakobsen, Kaj B},
  journal={Microwave and optical technology letters},
  volume={48},
  number={5},
  pages={832--834},
  year={2006},
  publisher={Wiley Online Library}
}

@article{gilbert1955optimum,
  title={Optimum design of directive antenna arrays subject to random variations},
  author={Gilbert, EN and Morgan, SP},
  journal={Bell System Technical Journal},
  volume={34},
  number={3},
  pages={637--663},
  year={1955},
  publisher={Wiley Online Library}
}

@inproceedings{yen2025practicality,
  title={On The Practicality of Superdirectivity in the {HF} Band},
  author={Yen, S and Wawrzynek, E and Filipovic, D},
  booktitle={Allerton Antennas Application Symposium},
  volume={},
  number={1},
  pages={},
  year={2025}
}

@article{lo1966optimization,
  title={Optimization of directivity ratio of an arbitrary and signal-to-noise antenna array},
  author={Lo, Y and Lee, S and Lee, Q},
  journal={Proc. IEEE},
  volume={54},
  number={8},
  pages={1033--1045},
  year={1966}
}

@ARTICLE{1956_Uzsoky_APASH,
  author = {Uzsoky, M. and Solym\'{a}r, L.},
  title = {Theory of Super-Directive Linear Arrays},
  journal = {Acta Physica Academiae Scientiarum Hungaricae},
  year = {1956},
  volume = {6},
  pages = {185--205},
  number = {2},
  month = {Dec.},
  doi = {10.1007/BF03157322},
}

@article{1981_Hansen_PIEEE,
  title = {Fundamental limitations in antennas},
  volume = {69},
  DOI = {10.1109/proc.1981.11950},
  number = {2},
  journal = {Proceedings of the IEEE},
  author = {Hansen,  R.C.},
  year = {1981},
  pages = {170–182}
}

@Book{POZAR,
  author    = {Pozar, D. M.},
  title     = {Microwave Enginnering},
  year      = {2011},
  edition   = {4th},
  publisher = {Wiley},
}

\end{document}